\documentclass[aip,reprint]{revtex4-1}
\usepackage{graphicx}
\usepackage{ulem}
\draft 

\begin{document}


\title{Relaxation of a single defect spin by the low-frequency gyrotropic mode of a magnetic vortex} 



\author{J. Trimble}
\affiliation{Department of Physics, Case Western Reserve University, Cleveland, Ohio 44106, USA.}

\author{B. Gould}
\affiliation{Department of Physics, Case Western Reserve University, Cleveland, Ohio 44106, USA.}

\author{F. J. Heremans}
\affiliation{Pritzker School of Molecular Engineering, University of Chicago, Chicago, IL 60637, USA.}
\affiliation{Center for Molecular Engineering and Materials Science Division, Argonne National  Laboratory, Lemont, IL 60439, USA}

\author{S.~S.-L.~Zhang}
\affiliation{Department of Physics, Case Western Reserve University, Cleveland, Ohio 44106, USA.}

\author{D. D. Awschalom}
\affiliation{Pritzker School of Molecular Engineering, University of Chicago, Chicago, IL 60637, USA.}
\affiliation{Center for Molecular Engineering and Materials Science Division, Argonne National  Laboratory, Lemont, IL 60439, USA}

\author{J. Berezovsky}
\email{jab298@case.edu}
\affiliation{Department of Physics, Case Western Reserve University, Cleveland, Ohio 44106, USA.}




\date{\today}

\begin{abstract}
We excite the gyrotropic mode of a magnetic vortex and observe the resulting effect on the spin state of a nearby nitrogen-vacancy (NV) defect in diamond. Thin permalloy disks fabricated on a diamond sample are magnetized in a vortex state in which the magnetization curls around a central core. The magnetization dynamics of this configuration are described by a discrete spectrum of confined magnon modes, as well as a low-frequency gyrotropic mode in which the vortex core precesses about its equilibrium position. Despite the spin transition frequencies being far-detuned from the modes of the ferromagnet, we observe enhanced relaxation of the NV spin when driving the gyrotropic mode. Moreover, we map the spatial dependence of the interaction between the vortex and the spin by translating the vortex core within the disk with an applied magnetic field, resulting in steplike motion as the vortex is pinned and de-pinned. The strong spin relaxation is observed when the vortex core is within approximately 250 nm of the NV center defect. We attribute this effect to the higher frequencies in the spectrum of the magnetic fringe field arising from the soliton-like nature of the gyrotropic mode when driven with sufficiently large amplitude.    
\end{abstract}

\pacs{}

\maketitle 

\section{Introduction}

The interaction between defect spin qubits and magnons in a magnetic material offers opportunities both for control and coupling of qubit states \cite{Trifunovic2013,Wolf2016a,Badea2016,Andrich2017a,Wolf2017,Lai2018,Candido2020,Fukami2021}, and for use of the spin qubit as a probe to study magnon phenomena \cite{VanDerSar2015,Wolfe2015,Page2016,Du2017,Lee-Wong2020,Lee-Wong2020a,McCullian2020,McCullian2020a,Liu}. In the prototypical system, negatively charged nitrogen vacancy (NV) defects in diamond are placed in proximity to a ferromagnetic  structure. An ac magnetic field is then used to drive dynamics of the NV spin, the magnon modes of the ferromagnet, or both. Meanwhile, the NV spin state can be initialized and monitored optically. 

In previous studies, several mechanisms for interaction have been observed. Simultaneous resonant driving of a magnon mode and the NV spin results in enhanced driving of the spin transition. But excitation of magnons far detuned from the spin transitions can also affect the NV spin. Through multiple-magnon processes, the overall chemical potential of the magnon bath is raised resulting in increased population of the magnon bath across the entire magnon spectrum \cite{Du2017}. These incoherent magnons at the NV spin transition frequency are a source of magnetic field noise that yield enhanced relaxation of the NV spin. In experiments with effectively continuous magnon spectra, driving the zero-wave-vector ferromagnetic resonance mode at frequencies lower than an NV spin transition excites magnons in the continuous spectrum of higher wave-vector magnons, some of which overlap the NV spin transition frequencies. In a geometrically confined system, the magnon spectrum becomes discrete, and enhanced spin relaxation is only observed when a magnon mode coincides both in frequency with an NV spin transition and spatially with the NV defect. This has previously been observed, for example, in a uniformly-magnetized 6-$\mu$m-diameter permalloy disk \cite{VanDerSar2015}, with discrete magnon modes in a range near the NV spin ground state transitions $\sim 3$~GHz.     

In this experiment, we study the interaction of an NV spin with a smaller 2-$\mu$m-diameter permalloy (Ni$_{0.81}$Fe$_{0.19}$) disk in a vortex magnetization state. A vortex state forms in a thin, soft, micron-scale ferromagnetic disk with a thickness sufficiently smaller than the diameter and with negligible anisotropy \cite{Cowburn1999,Shinjo2000}. The magnetization of a vortex state curls around the center of the disk, in-plane and tangential to the disk boundary, except at the core of the disk where the magnetization orients out-of-plane (see inset to Fig.~\ref{fig:setup}c.) The central region with out-of-plane magnetization is the vortex core, with a half width of $\sim 10$ nm, which is set by the exchange length \cite{Hollinger2003}. Due to the out of plane magnetization, the vortex core produces a strong dipole-like magnetic fringe field. The vortex core can be displaced across the disk by an external in-plane magnetic field $\vec{B} = (B_x,B_y)$ with the vortex core position $\vec{x}=(x,y) = c \chi_0 (B_y,-B_x)$ moving perpendicular to the applied field, with direction set by the vortex circulation direction $c=\pm 1$. In similar disks, we have previously measured $\chi_0 \approx 60$~nm/mT \cite{Trimble2021}.

As compared to the approximate uniform magnetization of the larger disk in Ref. \cite{VanDerSar2015}, these vortex states display more strongly confined magnon modes at higher frequencies ($>4$~GHz), and also a soliton-like gyrotropic mode of the vortex core at low frequency $f_g = 0.15$~GHz, as shown in Fig.~\ref{fig:setup}(c) \cite{Guslienko2005,Compton2006a}. The higher frequency magnon modes are quantized into azimuthal and radial modes, yielding nodes circulating about the vortex core, or at radial distances from the vortex core, respectively. The gyrotropic mode is characterized by uniform motion of the vortex core orbiting about its equilibrium position at frequency $f_g$. At low magnetic fields, the NV spin transitions occur near frequencies $f_0= 2.87$~GHz and $f_{ex} = 1.43$~GHz in the ground and first excited orbital states, respectively. Both $f_0$ and $f_{ex}$ lie well within the gap between the gyrotropic mode and the confined magnon modes. Naively, one would expect that driving any of the magnon modes in this system would not lead to NV spin relaxation, because there are no magnon modes resonant with NV spin transitions. However, we find that excitation of the gyrotropic mode does in fact enhance NV spin relaxation when the vortex core is in proximity to the NV defect. We attribute this to the soliton-like nature of the gyrotropic mode, in which many higher frequencies are present in its spectrum.

\section{Methods}
\begin{figure}[h]
 \includegraphics{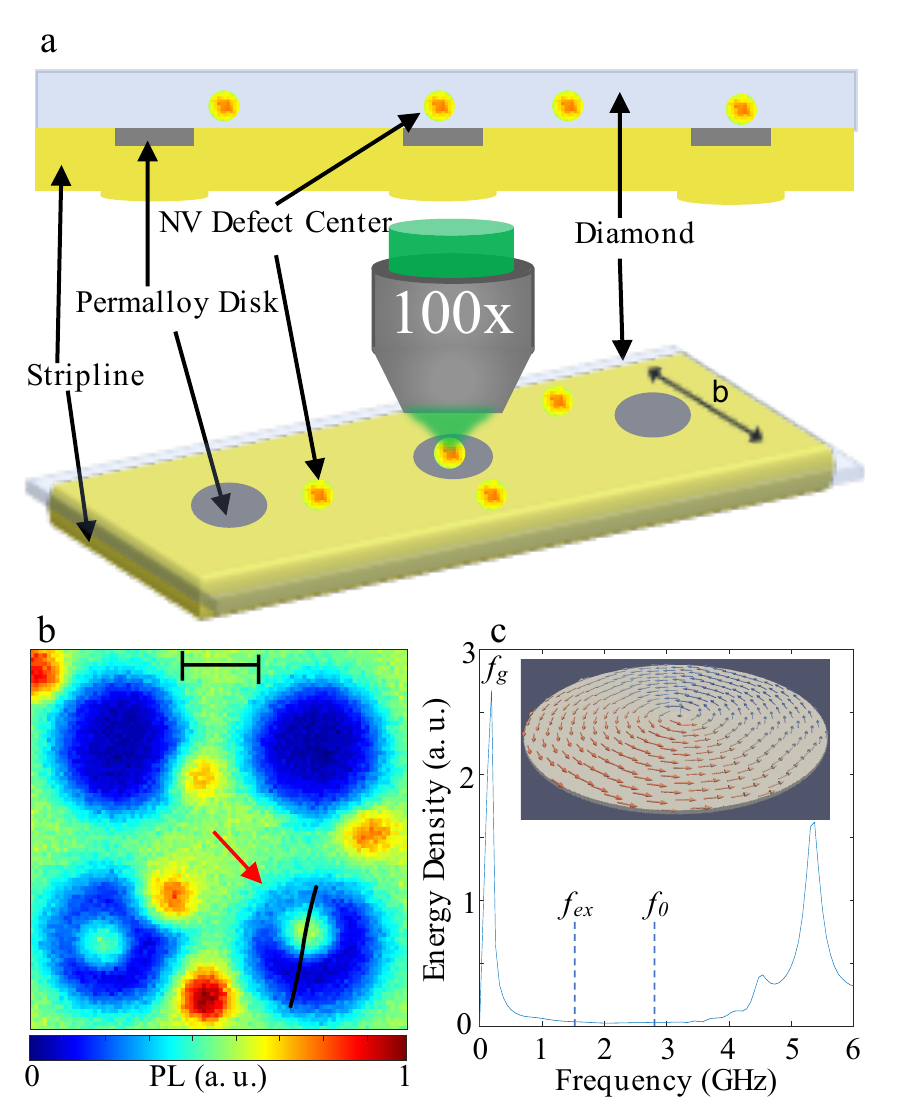}
 \caption{(a) Sample schematic. (b) Scanning PL image of four disks (blue) with several nearby NV defects (bright spots). The disk/NV pair used here is indicated by the red arrow. The calculated path of the vortex core over the range of $B_0 = -12$ to 12 mT is shown by the black line. Scale bar $= 1 \mu$m. (c) Simulated modes of the magnetic vortex, as excited by a weak broadband pulse. The zero-field splitting $f_0$ and $f_{ex}$ of the NV spin states in the ground and excited orbital states are indicated, in the gap between the gyrotropic mode and the confined magnon modes.\label{fig:setup}}
\end{figure}

A layer of isotopically pure $^{12}$C was grown on an electronic-grade single crystal diamond (Element Six) via plasma enhanced chemical vapor deposition.  $^{15}$N$_2$ gas was introduced into the growth process at the appropriate time to achieve a $15$ nm deep delta-doped layer of nitrogen.  A subsequent electron irradiation (2 MeV, 1e14 dose) was used to create the vacancies followed by annealing of the sample at $850$ $^{\circ}$C under forming gas (H$_2$/Ar) for 2 hours to form the NV centers.  A triacid (HClO$_4$:HNO$_3$:H$_2$SO$_4$) clean was then used to remove residual contaminants from the sample surface.  Details about the processing and growth of the NV defect center containing diamond film can be found in Refs. \cite{Ohno}, \cite{Ohno2014} and \cite{Mamin2014}.  An array of $2$-$\mu$m-diameter and $35$-nm-thick permalloy disks were then fabricated atop the NV-center-containing diamond film by electron beam lithography, electron beam evaporation and liftoff.  A subsequent photo-lithography process was used to pattern a $125$ nm thick gold co-planar waveguide (CPW) over the disk array.  Figure~\ref{fig:setup}(a) shows schematic of the sample and setup. A 100X oil immersion microscope with $1.25$ numerical aperture focuses 532-nm-wavelength excitation onto the sample from the back side of the diamond. Photoluminescence (PL) is collected back through the objective, and detected by a single photon counting module. Fig.~\ref{fig:setup}(b) shows a scanning PL image of four permalloy disks (blue), with several NV centers (bright spots) in the diamond $\approx 15$ nm above the disks.  The NV/disk pair studied in this paper is indicated by the red arrow.

The NV spin is initialized into the $m_s=0$ ground state with 532 nm laser excitation and read out at room temperature using confocal optically detected magnetic resonance (ODMR) spectroscopy. Transitions from $m_s = 0$ to the $m_s =\pm{1}$ sublevels cause a reduction in PL intensity. In the small magnetic fields $\vec{B}$ here, these transitions occur at $f_\pm \approx f_\mathrm{zfs} \pm (\gamma/2\pi) \vec{B}\cdot \hat{n}$, where $\hat{n}$ is the axis of the NV defect, and $f_\mathrm{zfs} = f_0$ or $f_{ex}$ for the orbital ground and excited states respectively. The gyromagnetic ratio $\gamma/2\pi = 28$~MHz/mT. To produce ODMR spectra, a microwave magnetic field $b$ is applied by driving the CPW at a frequency $f$ while the PL intensity is monitored. The value of $b$ at the position of the NV and vortex is calculated from the measured current in the CPW and the CPW geometry. Via an arrangement of permanent magnets on motorized stages, an in-plane static magnetic field $\vec{B}$ both splits the spin transitions, and also sets the position of the vortex core within the disk. We sweep the vortex along the line shown in black in Fig.~\ref{fig:setup}(b) by applying a field $\vec{B}=\vec{B}_\mathrm{off}+B_0 \hat{l}$, where $\vec{B}_\mathrm{off}$ is constant, and $B_0$ is swept along an axis $\hat{l}$. We collect ODMR spectra at different $B_0$, mapping out NV spin polarization as a function of the interaction with the magnetization of the disk and the applied field.  

 \begin{figure}[h]
 \centering
 \includegraphics{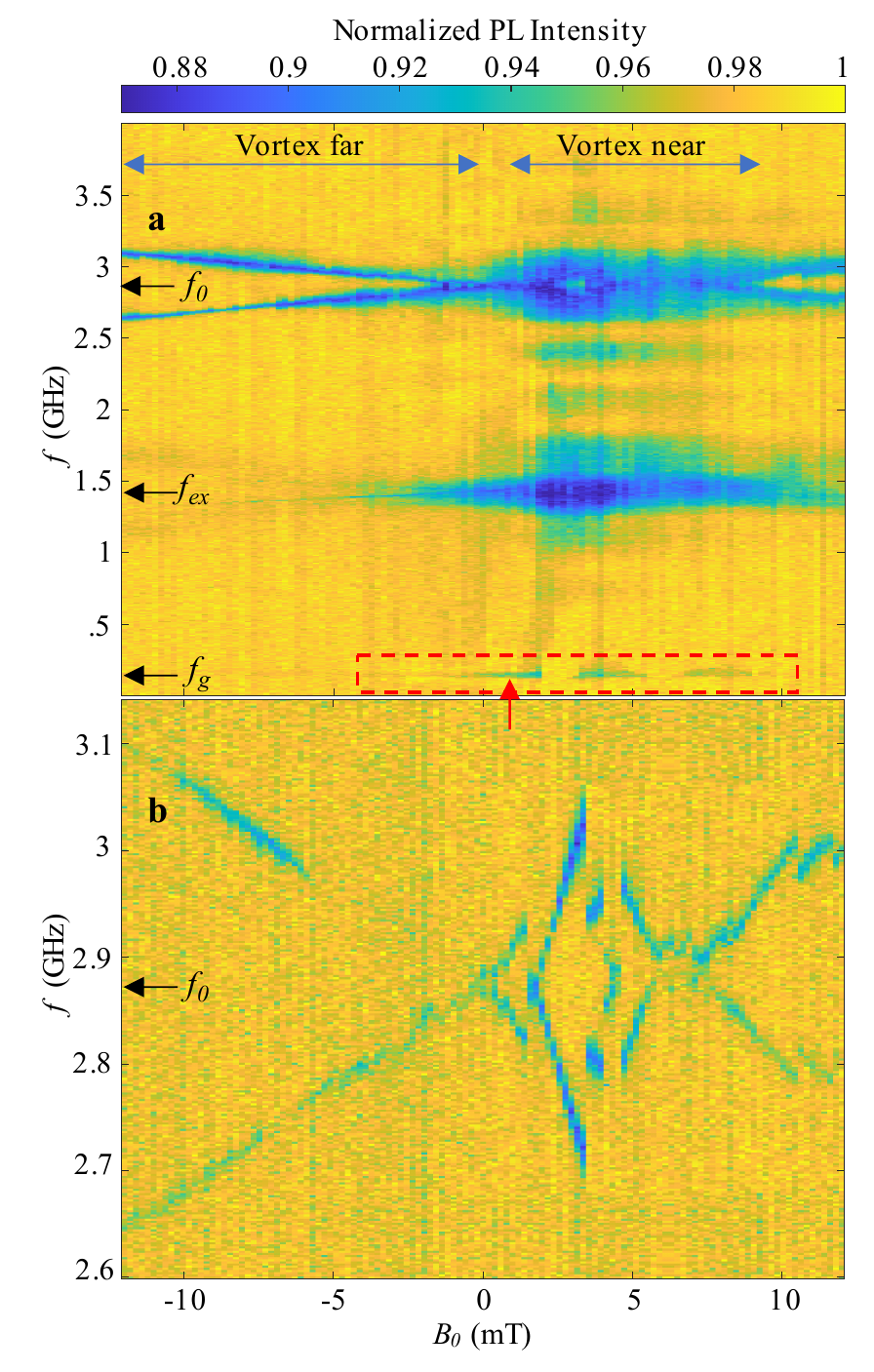}
 \caption{Optically detected magnetic resonance (ODMR) results. (a) ODMR spectra vs. $B_0$ with $b = 0.5$~mTrms. $B_0$ both splits the spin transitions, and translates the vortex across the disk. Regions where the vortex core is near and far from the NV are indicated. The features at $f_g$ are highlighted by the red dashed box, with the value of $B_0 = 1$~mT used in Fig.~\ref{fig:powerdep}(a) indicated by the red arrow. (b) ODMR scan at reduced $b=0.035$~mTrms in a range of $f$ near $f_0$, more clearly showing the ground state spin transitions.\label{fig:2dodmr}}
\end{figure}
\section{Results}

Figure \ref{fig:2dodmr}(a) shows the ODMR spectrum vs. $f$ and $B_0$ of a single NV defect center over a disk shown in Fig. \ref{fig:setup}(b). At $B_0<0$, the splitting of the NV center ground and excited states, centered at $f_0$ and $f_{ex}$, show the expected linear Zeeman splitting due to the applied magnetic field. At positive fields $B_0 = 0$ to $8$ mT, the vortex core passes near the NV and we observe a deviation from the Zeeman splitting due to the applied field. In this region, the NV experiences significant magnetic field from the vortex core. The largest effects occur when the NV-vortex core distance is small, around $B_0 = 2$ to $5$ mT. In addition to providing additional splitting due to the static magnetic fringe field, the vortex core field also causes additional broadening of the transitions.  As shown in Ref.~\cite{Wolf2016a}, When $f$ is resonant with a spin transition, even though $f$ is far off-resonance from any vortex modes, the small off-resonant motion of the vortex core is sufficient to substantially enhance the applied microwave driving.

In order to better see the effect of the vortex on the NV ground state transition frequencies, we perform an ODMR scan reducing the microwave field to $b=0.035$~mTrms from the $b = 0.5$~mTrms used above, and finer frequency step increment. Figure \ref{fig:2dodmr}(b) shows the ODMR spectrum of the same single NV defect in Fig. \ref{fig:2dodmr}(a), over the same range of $B_0$, but in this case over a narrower frequency range centered at $f_0$. The resonances in the region of strong NV-vortex interaction are now significantly narrower. Several discontinuities are observed in the splitting at positive $B_0$. These occur due to pinning and depinning events when the vortex becomes trapped at defects in the permalloy \cite{Badea2015b,Trimble2021}. The data in Fig 2(b) more clearly shows the additional splitting of the NV spin states when the vortex core is near the NV, with the maximum occuring around $B_0 = 4$ mT.

Beyond the ground and excited state NV transitions, however, additional features appear in the spectra in Fig.~\ref{fig:2dodmr}(a). In this work, we focus on the feature at $f = 0.15$~GHz highlighted by the red dashed box where intermittent dips in PL are observed when the vortex is close to the NV. This feature occurs at the expected frequency $f_g$ of the vortex gyrotropic mode,  suggesting that excitation of the vortex gyrotropic mode is affecting the NV spin in a way that causes ODMR contrast. The cause of this dip in PL contrast at $f_g$ is the focus of this paper. The origin of the several other broad features at higher frequencies in the region of strong vortex-NV interaction will not be addressed here.

 \begin{figure}[h]
 \centering
 \includegraphics[width=\linewidth]{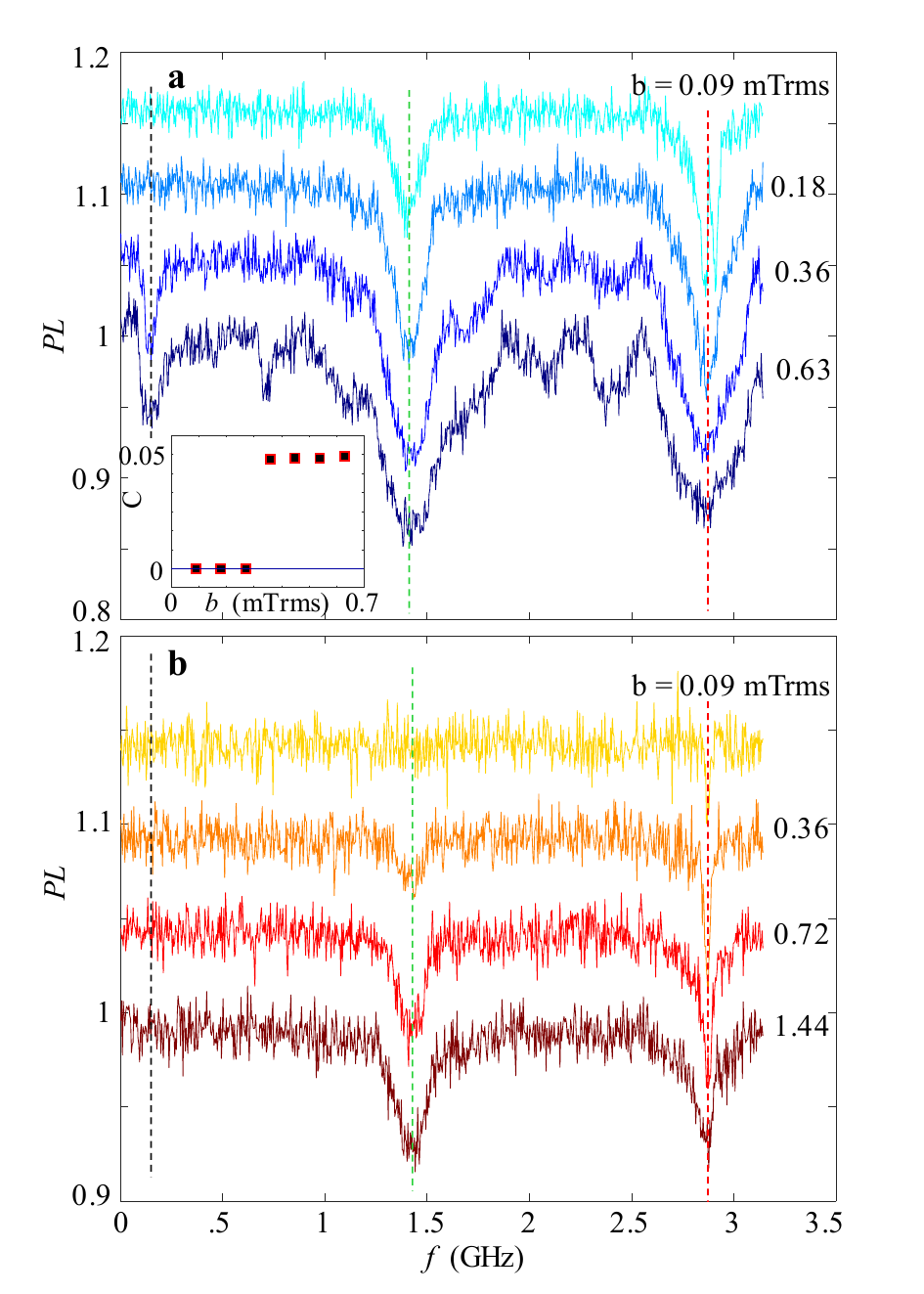}
 \caption{(a) ODMR spectra from the NV defect indicated in Fig.~\ref{fig:setup}(b) at fixed $B_0=1$~mT at increasing values of $b$. Frequencies $f_0$, $f_{ex}$, and $f_g$ indicated by red, green, and black dashed lines respectively. Inset: ODMR contrast $C$ at $f_g$ vs. $b$. (b) Similar to (a), but for an NV not near a vortex. \label{fig:powerdep}}
\end{figure}

Figure \ref{fig:powerdep}(a) shows a series of ODMR spectra with increasing values of $b$. These scans are performed with fixed $B_0=1$~mT, as indicated by the red arrow in Fig.~\ref{fig:2dodmr}. The expected dips in contrast near $f_0$ (red dashed line) and $f_{ex}$ (green dashed line) are observed, with increasing power broadening as $b$ is increased. The data show that at $b < 0.18$ mTrms the dip in contrast at the gyrotropic frequency (see black dashed line) is not observed, but at $b \geq  0.36$ mTrms the dip emerges. Further, the PL dip has the same magnitude at all powers tested from $b = 0.36$ to $0.63$ mTrms. The inset graph in Fig. \ref{fig:powerdep}(a) shows the $b$-dependence of the ODMR contrast $C$ at $f_g$. ($C$ is defined as the fractional reduction of PL intensity.) We attribute the sudden step in $C$ to vortex pinning. When the vortex core is pinned at a defect, the amplitude of the gyrotropic mode is significantly reduced, and the frequency may be shifted \cite{Compton2010a}. At sufficiently large $b$, the vortex can be unpinned and undergo larger amplitude motion. The fact that the scale of $b$ required to achieve nonzero $C$ is similar to the field scale between pinning jumps in Fig.~\ref{fig:2dodmr}(b) lends credence to this explanation. 

In order to verify that the dip in contrast at $f_g$ observed in Fig. \ref{fig:2dodmr}(a) and Fig. \ref{fig:powerdep}(a) is due to the excitation of the vortex gyrotropic mode, we investigated an NV defect that is not on or near a disk, so the NV only is affected by the applied static and microwave fields. Figure \ref{fig:powerdep}(b) shows line scans at $b = 0.09$  to $1.44$ mTrms. The only dips in contrast observed in these data are those near $f_0$ and $f_{ex}$ as expected. Note that the data in Fig.~\ref{fig:powerdep}(b) go to higher values of $b$ than in Fig.~\ref{fig:powerdep}(a). Despite the higher $b$ in Fig.~\ref{fig:powerdep}(b), the resonances are less broad than in Fig.~\ref{fig:powerdep}(a) because of the vortex enhanced broadening, as described in Ref.~\cite{Wolf2016a}.

 \begin{figure}[h]
 \includegraphics[width=\linewidth]{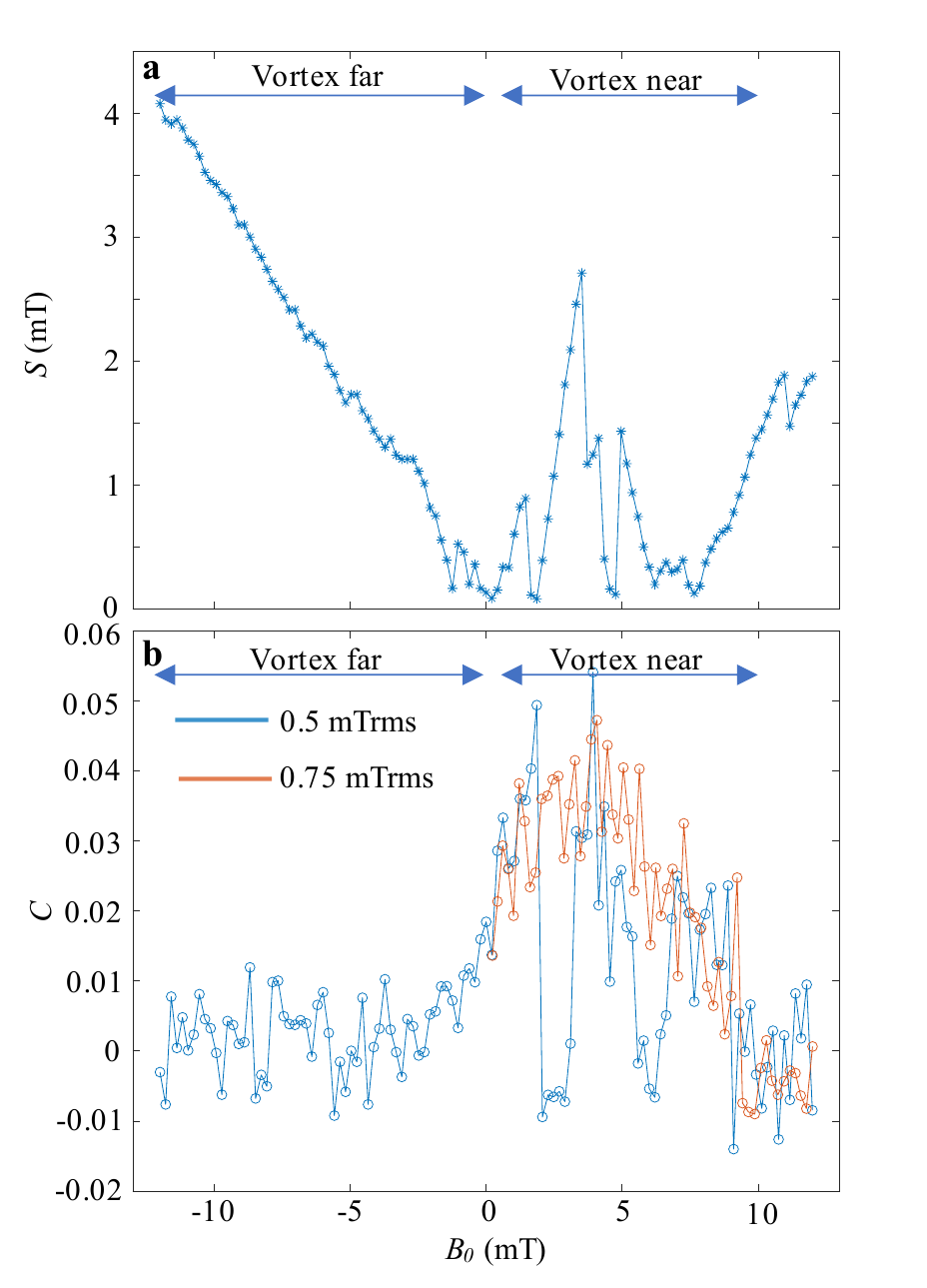}
 \caption{(a) Half splitting $S$ between $f_+$ and $f_-$ vs. $B_0$ for the ground state transition, extracted from the data in Fig.~\ref{fig:2dodmr} near $f_0$. (b) ODMR contrast $C$ at $f_g$ vs. $B_0$ at two values of $b$.\label{fig:contrast}}
\end{figure}

Fig.~\ref{fig:contrast} compares the effect of the static vortex core on NV spin splitting and the driven vortex core on the ODMR contrast. Fig.~\ref{fig:contrast}(a) shows the half splitting $S = \pi|f_+ - f_-|/\gamma$ of the ground state transitions about $f_0$ and Fig.~\ref{fig:contrast}(b) shows the ODMR contrast at $f=f_g$, both extracted from Fig.~\ref{fig:2dodmr}(a) and (b). The field $B_0$ both splits the spin transitions, and translates the vortex core across the disk, with the smallest separation between the vortex core and the NV occurring around $B_0 = 4$~mT. In Fig.~\ref{fig:contrast}(a), when the vortex core is far from the NV, we observe the continuous, linear Zeeman splitting as expected. From $B_0 = 0$ to $7.5$ mT the vortex passes near the NV and we observe strong and non-continuous splitting as the dipole-like fringe field of the vortex core passes near the NV in a series of pinned configurations. The ODMR contrast $C$ in Fig.~\ref{fig:contrast}(b) shows a peak in the same range as the enhanced $S$ in Fig.~\ref{fig:contrast}(a), providing further evidence that both effects depend on the proximity of the NV and vortex core. The blue data plotted in Fig.~\ref{fig:contrast}(b) are measured with $b = 0.5$ mTrms, as in Fig.~\ref{fig:2dodmr}(a), in which intermittent suppression of $C$ is seen, for example at $B_0 = 2$~mT. The red data in Fig.~\ref{fig:contrast}(b) are measured with an increased $b = 0.75$ mTrms (and a limited range of $B_0 > 0$.) Here, we see a similar envelope as at lower $b$, but without the jumps to $C=0$. This is consistent with the picture of the vortex core translating through different pinning configurations. When pinned, a certain threshold in $b$ is required to excite large-amplitude gyrotropic motion. Such a threshold at $b \approx 0.3$~mTrms was observed in the inset to Fig.~\ref{fig:powerdep}(a) at $B_0 =1$~mT. In Fig.~\ref{fig:contrast}(b), both values of $b$ are already above the threshold at $B_0 = 1$~mT. Though there is no precise correlation between the jumps in $S$ and the jumps in $C$, it is notable that the large dip in $C$ near $B_0 = 2$~mT coincides with the largest slope in $S$ vs. $B_0$ in Fig.~\ref{fig:contrast}(a). The large slope in $S$ may be caused by a strongly pinned vortex core deforming without translating \cite{Badea2015}. 

To understand how the gyrotropic mode of the vortex influences the NV spin, we performed simulations using the object-oriented micromagnetic framework \cite{OOMMF}. We simulate a 2-$\mu$m-diameter, $40$-nm-thick permalloy disk, with saturation magnetization $M_S=8.1 \cdot 10^5$~A/m and exchange stiffness $A=1.05 \cdot 10^{-11}$~J/m, on a rectangular mesh with cell size $(x \times y \times z) = (5 \times 5 \times 20)$~nm$^3$. The spectrum in Fig.~\ref{fig:setup}(c) was obtained by relaxing the magnetization at zero applied field, and applying a weak broadband pulse of the form $b_x = b_0 \sin{(2\pi f_c(t-t_0))}/(2\pi f_c(t-t_0))$ to excite the magnon modes~\cite{Lupo2015}. In this simulation, we let $b_0 = 1$~mT, $t_0 = 0.4$~ns, and the pulse contains frequency up to $f_c = 20$~GHz. The fast Fourier transform of the energy density vs. time in each cell in the disk was computed, and the magnitude in one representative cell is shown in Fig.~\ref{fig:setup}(c). 
We see good agreement between the experimental $f_g = 0.15$~GHz and the simulated $f_g = 0.165$~GHz, with the difference likely due to slightly different values of thickness, $M_s$, or $A$. As noted above, for this weak excitation there is no spectral overlap with the NV spin transitions. Indeed, in the data, we have observed a threshold in $b$ below which excitation in the range of $f_g$ does not affect the ODMR contrast. 

 \begin{figure}[h]
 \includegraphics[width=\linewidth]{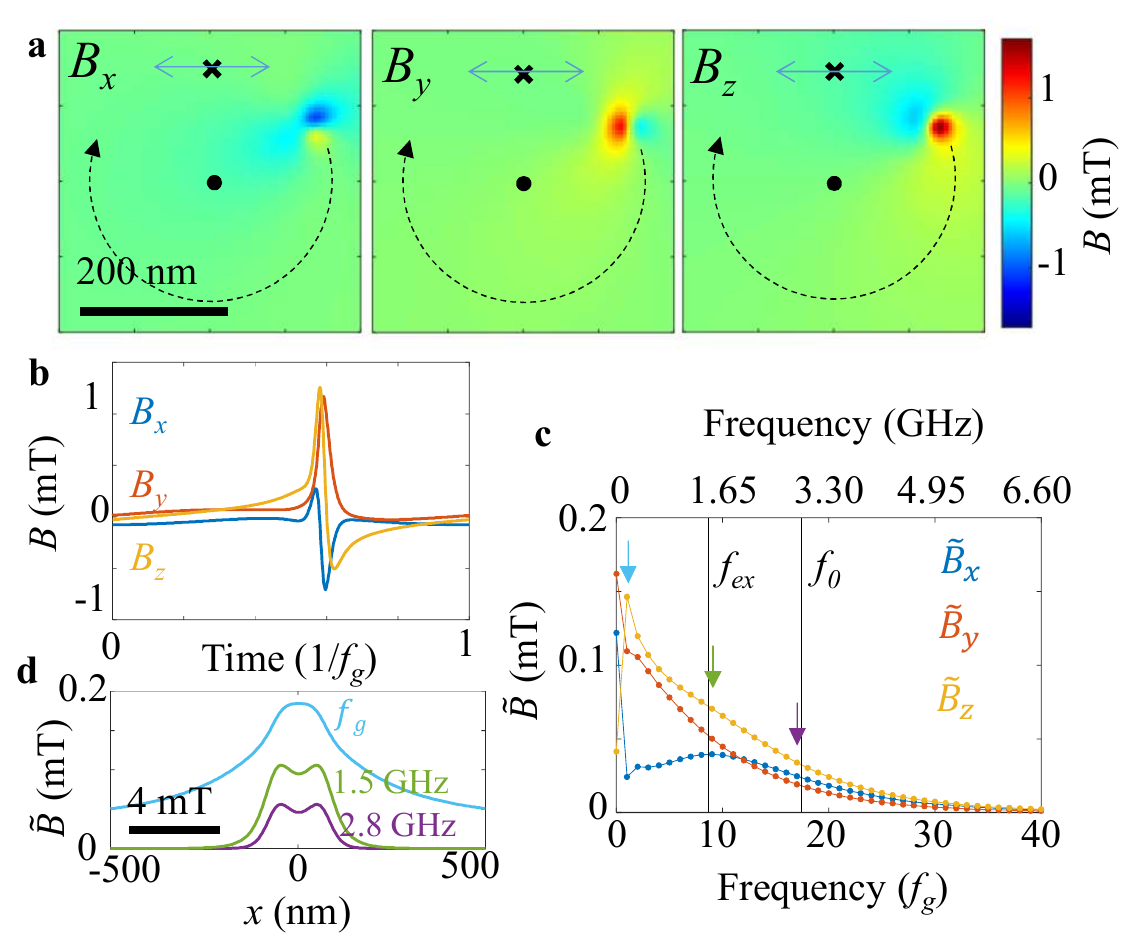}
 \caption{(a) Simulated fringe field components $B_x$, $B_y$, and $B_z$ in the cell above the disk as the vortex core is undergoing gyrotropic motion as indicated by the dashed circular arrows. (b) Magnetic field components at the point labeled `$\times$' in (a) over one period of the gyrotropic motion. (c) Fast Fourier transform amplitudes $\tilde{B}_x$, $\tilde{B}_y$, $\tilde{B}_z$ of the field components vs. time in (b), showing significant spectral overlap with $f_{ex}$ and $f_0$. (d) Amplitude $\tilde{B} = (\tilde{B}_x^2 + \tilde{B}_y^2 + \tilde{B}_z^2)^{1/2}$ at $f=f_g, 1.5$~GHz and 2.8 GHz, at a point translated a distance $x$ away from the `$\times$' in (a), as indicated by the double ended arrows in (a). The scale bar converts $x$ to in-plane field via $\chi_0$. \label{fig:sim}}
\end{figure}

The key fact in explaining the interaction between the gyrotropic mode and the NV spin transitions is that the spectrum of the fringe field generated by the gyrotropic mode is highly nonlinear with amplitude. This arises from the soliton-like nature of the gyrotropic mode, where the vortex core with radius $r_c \approx 10$~nm undergoes uniform displacement about the equilibrium position. We can estimate the typical radius of the gyrotropic orbit as $r_g \approx Q b \chi_0$, where $Q$ is the quality factor of the gyrotropic mode. From time-resolved measurements of the gyrotropic mode on similar disks~\cite{mehrnia2019three}, we estimate $Q\approx 5$. For a typical $b = 0.5$~mTrms, this yields $r_g \approx 150$~nm. With $r_g \gg r_c$, the local magnetization, and resulting properties such as energy density or fringe field, do not just vary sinusoidally at $f_g$ but also have significant higher frequency components as well.

To simulate the driven gyrotropic mode with large amplitude, we obtain a snapshot of the vortex undergoing gyrotropic motion with radius $r_g$, then rotate that magnetization about the center of the disk at frequency $f_g$. The snapshot of the vortex with the core at $r_g$ is obtained by first relaxing the vortex state at a relatively large field $B_x = 7$~mT to translate the vortex core to a radius $r>r_g$. The system then evolves over time at $B=0$ with a realistic Gilbert damping $\alpha = 0.008$, resulting in the vortex core orbiting around, and relaxing back to, the origin. When, during this trajectory, the vortex core has relaxed to $r=r_g$, we capture a snapshot of the magnetization and rotate about the disk center to simulate driven gyrotropic motion at constant $r_g$. This simulation method is more efficient than directly simulating a sinusoidal driving field until reaching steady state. The components of the dipole-like fringe field in the cell above the disk at one snapshot in time are shown in Fig.~\ref{fig:sim}(a), with the subsequent trajectory of the vortex core illustrated by the dashed circle.

If the NV is at the position indicated by the `$\times$' in Fig.~\ref{fig:sim}(a), it will experience sharp spikes in magnetic field with period $1/f_g$ each time the vortex core sweeps past. The fringe field at this position as shown in Fig.~\ref{fig:sim}(b) over one period. Note that the simulated amplitude of these pulses $\approx 1$~mT, is if anything an underestimate of the fringe field that may be generated by the dynamic vortex core: the splitting $S$ shown in Fig.~\ref{fig:contrast} indicates that the vortex core generates fringe fields up to several mT.

The frequency spectra $\tilde{B_x}$, $\tilde{B_y}$, and $\tilde{B_z}$ of the three components of the pulse in Fig.~\ref{fig:sim}(b) are shown in Fig.~\ref{fig:sim}(c). The sharpness of the pulse in fringe field as the vortex core passes the NV yields a very broad frequency spectrum, with significant weight at both $f_{ex}$ and $f_0$. Therefore, the pulse can drive transitions between the $m_s=0$ and $m_s=\pm 1$ spin states in both the ground and excited states of the NV center. The magnitude of the field on resonance with a transition sets the Rabi frequency $f_R \approx \frac{\gamma}{4\pi}b_{\perp} \sim 1.4$~MHz, where $b_{\perp}$ is the component of the resonant field perpendicular to the NV axis. The relation is only approximate because the direction of the time-dependent field is not constant. Over the time span of one gyrotopic period $1/f_g$, we then expect a Rabi rotation $\Delta \theta = f_R/f_g  \sim 0.01$ for each pulse. 

The magnitude and sign of the Rabi rotation will depend on the phase of the spin state relative to the arrival of the pulse. Since $f_g$ is not necessarily commensurate with $f_0$ or $f_{ex}$, the phase is essentially random between successive pulses. Moreover, the phase at successive pulses is very sensitive to dephasing of the spin state. The arrival of pulses at a rate $f_g$ with random phase of the spin state results in a random walk between the $m_s=0$ and $m_s=\pm 1$ eigenstates with a step size on the order of $\Delta \theta$, and hence, enhanced spin relaxation due to accumulation of random Rabi rotations. Within the repolarization time $\tau$ of the NV spin, we expect the random walk to rotate the spin through an angle with standard deviation $\sigma_\theta \sim \sqrt{\tau f_g}\Delta \theta$, assuming the phase of the spin state is completely decorrelated between successive gyrotropic periods. For $\tau \sim 1~\mu$s, we estimate $\sigma_\theta \sim 0.1$ radians. This order of magnitude is consistent with the observed PL dip, which is clearly visible, yet smaller than the dip due to complete depolarization at $f_0$ and $f_{ex}$.

To compare the spatial dependence of this spin relaxation mechanism to the experimental observations, we plot $\tilde{B}=(\tilde{B}_x^2+\tilde{B}_y^2+\tilde{B}_z^2)^{1/2}$ in Fig.~\ref{fig:sim}(d) at $f_g$ and near $f_0$ and $f_{ex}$, with the NV translated in the x-direction relative to the vortex core equilibrium position, as indicated by the double arrow in Fig.~\ref{fig:sim}(a). The actual magnitude of spin relaxation is a more complicated function of the pulse profile and components, but $\tilde{B}$ serves a simple measure of the strength of the contributing components. The position $x=0$, at the `$\times$', is the minimum separation between the vortex core and the NV, with a large spectral weight at all three frequencies. Both near $f_{ex}$ and $f_0$, $\tilde{B}$ falls off rapidly as the separation between the vortex core and the NV increases. The particular width and lineshape of these curves depends on the precise position and orientation of the swept path with respect to the gyrotropic orbit, but in general shows high values when the vortex core is in close proximity to the NV, and falls off quickly as they move apart. This is in good agreement with the experimental result shown in Fig.~\ref{fig:contrast}(b).  The scale bar in Fig.~\ref{fig:sim}(d) coverts position to applied magnetic field via $\chi_0$. 

\section{Conclusions}

We have shown that the driven gyrotropic mode of a magnetic vortex can cause enhanced spin relaxation of a nearby NV spin. Despite the large detuning between the NV spin transitions and any of the discrete modes of the magnetic vortex, we measure enhanced spin relaxation when the vortex core is sufficiently close to the NV and sufficiently strongly driven. Unlike a wave with a well-defined wavelength and frequency, a soliton propagates as a localized wave packet with a broad spectrum. The vortex core differs from a soliton in that its form is not governed by a nonlinear dispersion, yet the resulting traveling wave is still described by a broad spectrum of wavelength and frequency components. Particularly for large amplitude motion, these higher frequency components yield interactions that are not apparent from the spectrum of modes under weak excitation. The type of coupling between the gyrotropic mode of a vortex and the NV spin described here may also occur in other magnetic systems coupled to spin qubits, including those based on dynamically driven domain walls or skyrmions. 


%
%

%

\begin{acknowledgments}
We thank Jonathan Karsch and Sean Sullivan for careful reading of the manuscript and Kenichi Ohno for help with diamond growth.  Diamond sample  preparations at Argonne National Lab were primarily supported by the U.S. Department of Energy, Office of Science, Basic Energy Sciences, Materials Sciences and Engineering Division (F.J.H and D.D.A.) with support from the U.S. Department of Energy, Office of Science, National Quantum Information Science Research Centers. Work by S.~S.-L.~Z. was supported by College of Arts and Sciences, Case Western Reserve University.
\end{acknowledgments}

\section*{Data availability}

The data that support the findings of this study are available from the corresponding author
upon reasonable request.

\providecommand{\noopsort}[1]{}\providecommand{\singleletter}[1]{#1}%

\end{document}